\documentclass[10pt,twocolumn,aps,pra,showpacs,amsmath,amssymb,floatfix]{revtex4-1}

\usepackage{graphicx}
\usepackage{siunitx}
\usepackage{braket}
\usepackage{hyperref}
\usepackage{amssymb}
\usepackage{amsmath}

\begin{document}

\title{All-optical measurement of Rydberg state lifetimes}

\author{Markus Mack} 
\email[]{markus.mack@uni-tuebingen.de}
\author{Jens Grimmel}
\author{Florian Karlewski} 
\author{L\H{o}rinc S\'{a}rk\'{a}ny}
\author{Helge Hattermann}
\author{J\'{o}zsef Fort\'{a}gh}
\email[]{fortagh@uni-tuebingen.de}
\affiliation{CQ Center for Collective Quantum Phenomena and their Applications, Physikalisches Institut, Eberhard-Karls-Universit\"at T\"ubingen, Auf der Morgenstelle 14, D-72076 T\"ubingen, Germany} 

\date{\today}
\begin{abstract}
	We have developed an all-optical method for measuring the lifetimes of $n S$ and $n D$ Rydberg states and demonstrate its capabilities with measurements on a dilute cloud of ultracold ${}^{87}$Rb atoms in a cryogenic environment. The method is based on the time-resolved observation of resonant light absorption by ground state atoms and selective transfer of Rydberg atoms into the ground state at varying delay times in order to reconstruct Rydberg decay curves. Our measurements of the ${}^{87}$Rb $30S_{1/2}$ state indicate an increase of the lifetime at lowered environment temperatures, as expected due to decreased black body radiation. For the $38D_{5/2}$ state with an attractive dipole-dipole interaction, ionization and lifetime reduction due to collisional effects are observed.
\end{abstract}


\maketitle

\section{Lifetimes of Rydberg atoms\label{sec:lifetimes}}

Rydberg atoms are promising for quantum information processing due to their strong and highly tunable interaction properties \cite{Saffman.2010}. High fidelity quantum gates and coherent state transfer between Rydberg and long living ground states have been proposed \cite{Jaksch.2000, Mueller.2014}. The fidelity of these operations is, however, fundamentally limited by the finite lifetime of Rydberg states \cite{Saffman.2010}. Besides the natural decay of the Rydberg excitation, blackbody radiation induced transitions \cite{Beterov.2009, Beterov.2009.Err}, collisions \cite{WalzFlannigan.2004}, and superradiance \cite{Wang.2007} may also limit the lifetime. The characterization of the Rydberg state decay is thus of significant interest.

For individual Rydberg atoms at an environment temperature $T=0$ the lifetime of an excited state is given by the inverse sum over all spontaneous decay rates into lower lying states \cite{Gallagher.1994}. Due to the highest energy difference, the lowest lying states contribute most to the decay. This is a limiting factor for calculations because the potentials for low-lying states can not be described as accurately as those of higher states, which become more and more hydrogen-like with increasing $n$ and $l$ quantum numbers. In a finite temperature environment, blackbody radiation (BBR) induced transitions occur. The strongest transitions are those to nearby dipole-allowed Rydberg states both above and below in energy. For a perfect Planck photon distribution and well-known temperature the corresponding rates can be calculated with high accuracy \cite{Beterov.2009, Beterov.2009.Err}. The experimental verification of BBR induced transition rates is possible not only through Rydberg lifetime measurements \cite{Spencer.1982} but also indirectly by e.g. measurements of Stark maps \cite{Grimmel.2015}, which depend on the same dipole matrix elements. Any incoherent re-population of the originally excited Rydberg state by multiple BBR transitions can be easily included into theoretical models, but is usually negligible in magnitude. Also, blackbody-induced ionization by transitions to continuum states can be taken into account \cite{Beterov.2009_2}.

Direct lifetime measurements at lowered environment temperatures, as well as measurements of temperature-dependent BBR transfer rates, have been conducted for Na atoms \cite{Spencer.1981, Spencer.1982}. The most accurate values for Rb Rydberg lifetimes to date have been measured in a room temperature environment, relying on the knowledge of BBR transition rates in order to extract zero-temperature natural lifetimes. Measurements of $nS$ and $nD$ states in the range of $n=\text{\numrange{27}{44}}$ were conducted by exciting Rydberg atoms from a cloud of ultracold atoms prepared in a magneto-optical trap (MOT), waiting some varying delay time, and then applying an electric field pulse while monitoring the time-dependent ionization signals (selective field ionization, SFI) \cite{Oliveira.2002, Nascimento.2006}. Due to the difficulty of accurately distinguishing between close lying Rydberg states which are populated because of BBR (see discussion \cite{Tate.2007, Caliri.2007}), the technique was improved in \cite{Branden.2010}. By adding a microwave transfer of the "target" Rydberg atoms to a higher-lying state which can be accurately discriminated, this potential source of systematic error was eliminated. The results of \cite{Branden.2010} generally agree with the previous work and cover Rb $nS$, $nP$, and $nD$ states in the range $28 \leq n \leq 45$. To our knowledge, neither lifetimes of Rb at lowered environment temperatures, nor any BBR transition rates, have yet been measured. Consequently, an experimental verification of the BBR rate calculations \cite{Beterov.2009} for Rb is still required.

In general, the lifetime of Rydberg atoms in ultracold gases is altered by several effects. Any electric fields lead to state mixing and ionization \cite{Gallagher.1994}. Collisions between atoms, as well as dipole-dipole and higher order interaction between Rydberg atoms, which may also lead to collisions \cite{Li.2005}, cause changes of the atomic states and ionization \cite{WalzFlannigan.2004, Zanon.2002}. Furthermore, depending on atomic density and cloud geometry, microwave superradiance is likely to occur, which can be triggered by blackbody radiation \cite{Wang.2007, Day.2008}. Due to such effects, the lifetimes of Rydberg atoms can differ greatly from the undisturbed values, as well as from one experiment to another. Therefore, the measurement of Rydberg lifetimes under the given conditions is necessary.

While the SFI methods mentioned above can be used when an electron/ion detector is present, an increasing number of recent cold atom experiments rely solely on optical detection, mostly employing electromagnetically induced transparency (EIT) instead \cite{Tauschinsky.2010, Guenter.2013, Maxwell.2014, Gorniaczyk.2014}. In order to enable the determination of Rydberg lifetimes in such systems, as well as in cases where a reduced complexity of the setup, compared to the SFI approach with the additional microwave, is desired, we developed a similarly powerful, all-optical method for measuring state-specific lifetimes. The approach is technologically simplified as the same lasers that are used for Rydberg excitation are employed for detection, requiring only an additional photodiode for the measurement of resonant absorption.

\begin{figure}
	\includegraphics{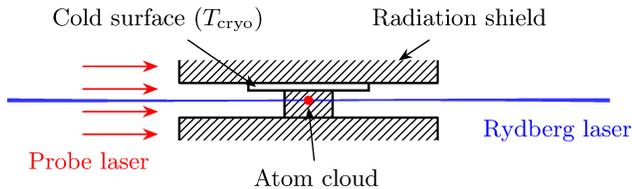}
	\caption{\label{fig:setup}(Color online) Cloud of ultracold ${}^{87}$Rb atoms (center) inside the radiation shield of a cryostat (hatched). The radiation shield is cylindrical in shape with a \SI{4.5}{\milli\meter} gap and \SI{35}{\milli\meter} diameter. While the cold surface is at $T_\mathrm{cryo}$, the temperature of the radiation shield is slightly higher. The outside temperature is assumed to be close to \SI{300}{\kelvin}.}
\end{figure}

We describe the optical lifetime measurement of Rydberg states in Sec. \ref{sec:method} and demonstrate its application in a setup with cold ${}^{87}$Rb atoms (see Fig. \ref{fig:setup}) at cryogenic environmental temperatures in Sec. \ref{sec:measurements}. Factors influencing the accuracy of the method are discussed in Sec. \ref{sec:discussion}.

\section{Optical lifetime measurement method \label{sec:method}}

\begin{figure}
	\includegraphics{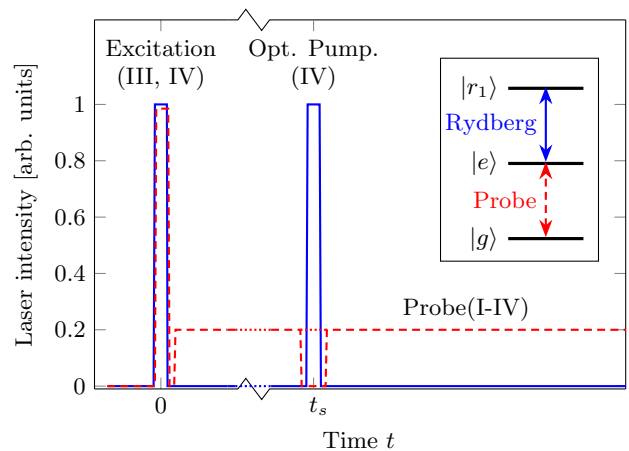}
	\caption{\label{fig:pulse_sequence}(Color online) Illustration of the pulse sequence for all steps I to IV used for the optical lifetime measurement, for the probe (dashed red lines) and Rydberg (solid blue lines) lasers, resonant to the $\ket{g}\leftrightarrow\ket{e}$ and $\ket{e}\leftrightarrow\ket{r_1}$ transitions. For the calibration (step I without and II with atoms), only the probe laser is turned on after $t=0$ for a duration of several hundred microseconds ($\gg$ Rydberg lifetime, only initial part shown in plot). Starting at step III, at $t=0$ atoms are excited to the Rydberg state $\ket{r_1}$ (Excitation). For step IV, which is repeated several times, the long probe pulse is interrupted at time $t_s$ for a short duration, during which an optical pumping pulse of the Rydberg laser is applied (Opt. Pump.). $t_s$ is varied with each repetition of step IV.}
\end{figure}

The optical measurements presented in this article rely on time-resolved resonant absorption detection in an effective three-level ladder-type system as shown in the inset of Fig. \ref{fig:pulse_sequence}, similar to the scheme in \cite{Karlewski.2015}. The transmission of a \emph{probe laser} pulse resonant to a closed cycling transition between a ground state $\ket{g}$ and a quickly decaying intermediate excited state $\ket{e}$ is monitored with a photodiode. The duration of this pulse should be several times the expected Rydberg lifetime. The lifetime of $\ket{e}$ must be shorter than the expected time resolution of the final Rydberg decay curves. For Rydberg excitation a \emph{Rydberg laser} resonant to the transition between $\ket{e}$ and the target Rydberg state $\ket{r_1}$ is used simultaneously with the probe laser. In principle, this configuration allows excitation by means of a STIRAP pulse \cite{Cubel.2005}.
The experimental sequence, aimed at measuring changes in the optical density due to the laser pulses, consists of several steps. In each step, a cloud of ultracold atoms is prepared, released from the trap, and after a given time of flight a series of laser pulses depending on the current step is applied, as shown in Fig. \ref{fig:pulse_sequence}.

As a baseline calibration, the probe laser light intensity signals without any atoms (step I, signal $I_0(t)$) and with ground state atoms (step II, $I_{\mathrm{g}}(t)$) are recorded, giving the time-dependent optical density (Lambert-Beer law):
\begin{equation}
\label{eqn:OD} \mathrm{OD}_{\mathrm{no exc.}}(t) = -\log(I_{\mathrm{g}}(t)/I_0(t)),
\end{equation}
In general, during the relevant time scales, the OD is proportional to the number of atoms in the volume of the cloud "seen" by the probe laser beam.

In step III, the Rydberg excitation at $t=0$ just before the start of the probe pulse is added. Due to excited atoms that are missing from the ground state, there will be increased transmission as compared to step II. Again, an optical density of ground state atoms $\mathrm{OD}_{\mathrm{w/exc.}}(t)$ can be calculated as in \eqref{eqn:OD}.
In combination with the result from step II, an additional quantity
\begin{equation}
\label{eqn:p_notg} p_{\mathrm{\neq g}}(t) = 1 - \mathrm{OD}_{\mathrm{w/exc.}}(t) / \mathrm{OD}_{\mathrm{no exc.}}(t)
\end{equation}
can be determined, which gives the number of atoms \emph{not} in the ground state due to the excitation pulse, normalized to the total number of atoms in the detection volume. The value of $p_{\mathrm{\neq g}}$ in the beginning is the fraction of atoms that have been excited to the Rydberg state, except for transitions to other states that have already happened due to BBR and possible superradiance, as was noted in \cite{Karlewski.2015}. The whole $p_{\mathrm{\neq g}}$ curve represents an effective decay of all directly and indirectly excited states, which is nearly, but generally not perfectly, exponential in shape because of the differing lifetimes of the constituent Rydberg states that become populated. Also, if ionizing effects played a role, the curve will not return to zero for long times, but converge towards a finite value. The resulting $p_{\mathrm{ion}} = p_{\mathrm{\neq g}}(t\rightarrow\infty)$ is a measure for the strength of any ionizing effects, if other mechanisms can be excluded that specifically remove Rydberg atoms, but not ground state atoms, from the detection volume, or, alternatively, transfer them into other stable states outside the probe transition. In general, the decay curve must consist of the (as yet unknown) parts
\begin{equation}
\label{eqn:decayparts} p_{\mathrm{\neq g}}(t) = p_{\mathrm{r_1}}(t) + p_{\mathrm{r_{\neq 1}}}(t) + p_{\mathrm{ion}}(t),
\end{equation}
i.e. the population of the originally excited Rydberg state, other Rydberg states, and the number of ionized atoms.

In order to separate the decay of the originally excited Rydberg state $\ket{r_1}$ from others that become populated, a state transfer similar to the microwave transfer in \cite{Branden.2010} is employed in step IV. However, instead of the additional microwave, the same Rydberg laser which was used for the excitation is used for a short optical pumping pulse at various times $t_s$ during the expected decay of $\ket{r_1}$. This pumps a fraction of atoms still in the $\ket{r_1}$ state down to the intermediate $\ket{e}$ state. During this pulse, the probe laser needs to be turned off to prevent any re-excitation of the Rydberg state. Because the lifetime of the intermediate state is short, the atoms pumped back in this fashion will reappear as ground state atoms as soon as the probe laser is turned on again after the optical pumping pulse. Following the same evaluation procedure as for step III, using \eqref{eqn:OD} and \eqref{eqn:p_notg}, decay curves $p_{\mathrm{\neq g}, s}(t)$ can be obtained. These must consist of the same parts with equal values as \eqref{eqn:decayparts}, except for a change in $p_{\mathrm{r_1}}(t)$ at $t \geq t_s$, which has been reduced by some fraction $\alpha$ by the optical pumping pulse due to the Rydberg population in $\ket{r_1}$ at time $t=t_s$, leaving 
\begin{equation}
p_{\mathrm{r_1}, s}(t_s) = (1-\alpha) p_{\mathrm{r_1}}(t_s).
\end{equation}
Thus, by subtracting the decay curves $p_{\mathrm{\neq g}, s}(t)$ from the curve without optical pumping $p_{\mathrm{\neq g}}(t)$ of step III, information about the original $\ket{r_1}$ population at time $t=t_s$ can be obtained:
\begin{equation}
\alpha p_{\mathrm{r}_1}(t_s) = p_{\mathrm{\neq g}}(t_s) - p_{\mathrm{\neq g}, s}(t_s).
\end{equation}
As long as the optical density of the Rydberg atoms for the Rydberg laser is small or the optical pumping is fast enough to transfer all of the $\ket{r_1}$ atoms ($\alpha=1$), $\alpha$ will be a constant fraction for each $t_s$. Repetition of step IV for different $t_s$ and evaluation of $\alpha p_{\mathrm{r_1}}(t_s)$ yields the decay of the population $p_{r_1}$, giving the lifetime of the Rydberg state $\ket{r_1}$. In case only spontaneous decay and BBR contribute to the lifetime, this will be an exponential decay with a decay parameter $\tau$ independent of $\alpha$.

\section{Conducted measurements and results \label{sec:measurements}}

We employed the method in a series of experiments in a setup where a cloud of \numrange{4e5}{8e5} ${}^{87}$Rb atoms with a temperature of \SI{1.5}{\micro\kelvin} is transferred into a gap of the radiation shield of a tunable temperature cryostat by means of optical tweezers (detailed in \cite{Cano.2011}). Details of the geometry are shown in Fig. \ref{fig:setup}.
For excitation and detection we used the ${}^{87}$Rb $5S_{1/2}(F{=}2) \leftrightarrow  5P_{3/2}(F{=}3)$ transition and a circular polarization, which is commonly used for imaging purposes. A repumping laser resonant to the $5S_{1/2}(F{=}1) \leftrightarrow 5P_{3/2}(F{=}2)$, which was needed for the MOT operation as well, was used to effectively keep the $F=1$ ground state unpopulated at all times. The Rydberg excitation to $\ket{r_1}$ is done by a pulse of the probe laser, with a higher intensity ($\approx \times 5$) than in actual probing, and simultaneously the Rydberg laser pulse resonant to the $5P_{3/2}(F{=}3) \leftrightarrow 30S_{1/2}$ or $38D_{5/2}$ state. The probe and Rydberg lasers are frequency stabilized to a frequency comb and arranged as described in \cite{Karlewski.2015}. The Rydberg laser was focused down to a size of \SI{\approx 100}{\micro\meter} with a total power of \SI{\approx 20}{\milli\watt}. The measurement sequence as described in Sec. \ref{sec:method} was conducted at a time of flight (TOF) of \num{12.0} and \SI{20.5}{\milli\second} in order to reduce the atom density to \SI{7+-3e9}{\per\centi\meter\cubed} and \SI{1.5+-0.5e9}{\per\centi\meter\cubed}, respectively. The setup did not allow for controlled compensation of stray electric fields (which have been investigated in detail elsewhere \cite{Hattermann.2012}), leaving residual fields of $\SI{6.6}{\volt\per\centi\meter}$ for the $30 S$ state measurements, and, after optimizing the cloud position, $\SI{0.5}{\volt\per\centi\meter}$ for those in the $38 D$ state.
Due to limited Rydberg laser power, pulse durations of \SI{1}{\micro\second} were chosen for both the excitation and optical pumping pulses. In order to reduce the statistical errors, mainly a result of photodiode and other technical noise, $\num{\geq 30}$ shots per step were averaged. The time resolution of the photodiode signal of each shot was digitally reduced to \SI{0.5}{\micro\second} by temporal averaging. Furthermore, because of atom number fluctuations, \eqref{eqn:OD} and \eqref{eqn:p_notg} were individually calculated for each single-shot measurement and corrected for atom number drifts. In order to decrease statistical noise, the results of shots belonging to the same step (and repetition number $s$ for step IV) were averaged.

\begin{figure}
	\includegraphics{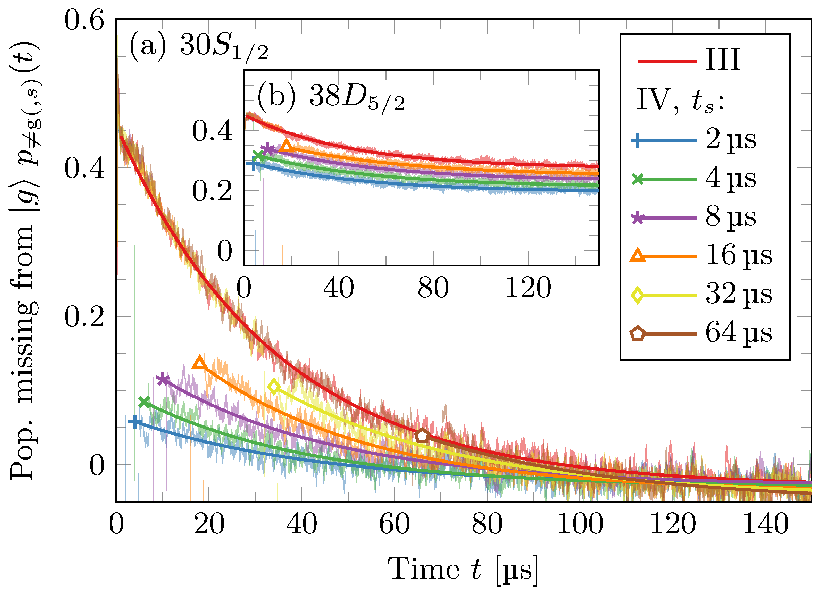}
	\caption{\label{fig:notg_decay}(Color online) (a) Time-dependent populations of states not contributing to the probe transition, $p_{\mathrm{\neq g}}$ (step III), and $p_{\mathrm{\neq g}, s}(t)$ (step IV with repetition number $s$ and corresponding optical pumping time $t_s$), representing an intermediate result of the evaluation of the measurement of the $30 S_{1/2}$ state at $T_\mathrm{cryo}=\SI{300}{\kelvin}$ and a TOF of $\SI{20.5}{\milli\second}$. Shaded areas show statistical errors of the signal at each instant of time. Solid lines show exponential fits corresponding to each measurement step. Markers show the values $p_{\mathrm{\neq g}, s}(t_s)$ obtained from the fit. (b) $38 D_{5/2}$ state at $T_\mathrm{cryo}=\SI{160}{\kelvin}$, showing only a fraction of atoms returning to the ground state depending on the optical pumping time $t_s$, converging towards values up to $p_{\mathrm{ion}} \approx 0.3$.}
\end{figure}

 Fig. \ref{fig:notg_decay} shows exemplary results for the number of atoms missing from the ground state, $p_{\mathrm{\neq g}}(t)$ and $p_{\mathrm{\neq g}, s}(t)$, where up to six values of $t_s$ have been used at increasing intervals. For both states, an approximation of the curves by exponential functions was sufficient, which were used for fitting in order to determine the points $p_{\mathrm{\neq g}, s}(t_s)$. While for $30 S_{1/2}$ all atoms eventually return to the ground state, for the $38 D_{5/2}$ state this is clearly not the case. This is most probably caused by ionizing collisions due to the known attractive dipole-dipole interaction for this state \cite{Reinhard.2007}. While more than half of the Rydberg atoms are apparently already lost at $t<\SI{2}{\micro\second}$, a small fraction was prevented from being ionized by the optical pumping pulses.

\begin{figure}
	\includegraphics{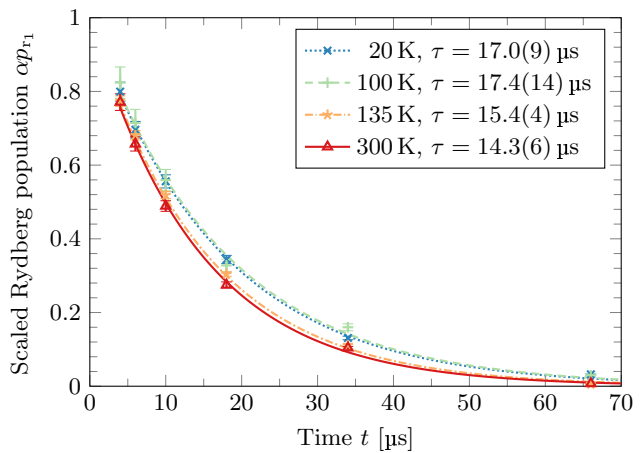}
	\caption{\label{fig:lifetime_30S}(Color online) Measured populations $\alpha p_{\mathrm{r_1}}(t_s)$ of the $30 S_{1/2}$ state, proportional to the decays of $\ket{r_1}$, at several cryostat temperatures in the range \SIrange{20}{300}{\kelvin}. The lines are exponential fits to the data, giving the decay parameters $\tau$ and their estimated error based on the fit (add $\SI{\pm 1.0}{\micro\second}$ systematic error common to all measurements for absolute uncertainty).}
\end{figure}

The final evaluation step yielding $\alpha p_{\mathrm{r}_1}(t_s)$ is shown in Fig. \ref{fig:lifetime_30S} for the state $\ket{r_1}=30S_{1/2}$, measured at cryostat temperatures in the range of \SIrange{20}{300}{\kelvin}, at a TOF of $\SI{20.5}{\milli\second}$, resulting in values $\tau{=}\SI{14.3+-0.6}{\micro\second}$ ($T_\mathrm{cryo}{=}\SI{300}{\kelvin}$) up to  $\tau{=}\SI{17.0+-0.8}{\micro\second}$ ($T_\mathrm{cryo}{=}\SI{20}{\kelvin}$) with an additional common systematic error of $\SI{\pm1.0}{\micro\second}$. Compared with the established measured value $\tau{=}\SI{14.5+-1.2}{\micro\second}$ ($30S_{1/2}, T{=}\SI{300}{\kelvin}$, Fig. 2 in \cite{Nascimento.2006}), our result is well compatible.

However, while in our measurements there is a clear tendency for increased lifetimes at lowered temperatures the effect is not as pronounced as expected from the calculated values from \cite{Beterov.2009}, giving $\tau{=}\SI{26.6}{\micro\second}$ at $T{=}\SI{0}{\kelvin}$. In order to check for any systematic dependencies on geometrical effects arising due to the falling and expanding cloud, as well as the presence of any density-dependent lifetime reducing effects, the measurements for $30S_{1/2}$ were repeated with a time of flight of $\SI{12}{\milli\second}$. This resulted in an increase of the atom density and the optical density by roughly a factor of 2. The resulting lifetimes for $30S_{1/2}$ were $\tau{=}\SI{14.1+-1.4}{\micro\second}$ ($T_\mathrm{cryo}{=}\SI{300}{\kelvin}$) up to $\tau{=}\SI{16.0+-0.7}{\micro\second}$ ($T_\mathrm{cryo}{=}\SI{20}{\kelvin}$), i.e. no discernible difference compared to the TOF $= \SI{20.5}{\milli\second}$ measurements. Therefore, we suspect insufficient shielding from outside thermal radiation in our setup as the main cause for the discrepancy in comparison to theory at low temperatures. The geometry as shown in Fig. \ref{fig:setup} leads to a solid angle of $4\pi\cdot\num{0.87}$ covered by cold surfaces. Also, the lower part of the radiation shield is not ideally thermally coupled to the cryostat and is estimated to be at least $\SI{20}{\kelvin}$ warmer than the upper surface at temperatures near $T_\mathrm{cryo}=\SI{4.2}{\kelvin}$, causing the effective temperature relevant for BBR to be significantly higher than the cryostat temperature. 

The lifetime of the $38D_{5/2}$ state, with a TOF of $\SI{20.5}{\milli\second}$, was measured to be $\SI{13+-4}{\micro\second}$ ($T_\mathrm{cryo}{=}\SI{160}{K}$) and $\SI{19+-3}{\micro\second}$ ($T_\mathrm{cryo}{=}\SI{20}{K}$), much lower than the reference value of $\tau{=}\SI{30(2)}{\micro\second}$ ($T{=}\SI{300}{\kelvin}$, Fig. 2 in \cite{Nascimento.2006}). This lifetime reduction comes as no surprise, due to the interaction-induced ionizing collisions already observed in Fig. \ref{fig:notg_decay}. 

\section{Accuracy considerations \label{sec:discussion}}

The time resolution of the method presented in Sec. \ref{sec:method} is limited by the length and timing accuracy of the excitation and optical pumping pulses, effectively adding uncertainty to times $t=0$ and $t_s$. For our $\SI{1}{\micro\second}$ pulses with a timing accuracy of $\SI{< 10}{\nano\second}$, this results in a systematic uncertainty of $\SI{\pm 1.0}{\micro\second}$ common to all measurements taken under the same excitation conditions, limiting their absolute, but not differential accuracy. Higher laser intensities, particularly of the Rydberg laser, would allow for shorter pulses. When choosing the sizes for both laser beams, geometry effects need to be considered: If there is any significant atomic motion due to time of flight or atomic temperature, the excitation volume should be smaller than the detection volume, in order to avoid any Rydberg atoms leaving the detection volume during the measurement time. However, the measured Rydberg signal will be lower for increased detection volumes, affecting the signal-to-noise ratio. Furthermore, prolonged acceleration of the atoms due to radiation pressure will lead to Doppler shifts which can lead to systematical errors, especially with regard to the narrow Rydberg transitions. This is particularly relevant for the probe laser, which must be well below the saturation intensity of the probe transition ($I_0 \lesssim \num{0.2} I_\mathrm{sat}$ in our experiments).

If, like in our experiments, the pulses are created using an acousto-optical modulator (AOM) in the laser beam path, the probe pulse will have a certain switching time of $\approx\SI{100}{\nano\second}$ and will show intensity drifts resulting from polarization drifts during some device-dependent warm-up time. As long as the signal remains proportional to the light level seen by the atoms, and the pulse shapes are well reproducible, these drifts cancel out when calculating the $\mathrm{OD}(t)$ terms.
The optical density itself will not be constant over the duration of the laser pulse, since the atomic cloud is expanding, as well as being accelerated downwards by gravity, however, these effects cancel out as well in $p_{\mathrm{\neq g}}(t)$.

The measured optical densities depend on the probe light polarization and Zeeman substates of the atoms. For low Rabi frequencies, optical pumping effects can become visible in the dynamic parts of the measured signals, particularly when turning the probe laser back on after the Rydberg laser pulses at times $t_s$, which in turn limits the accuracy of the determination of $p_{\mathrm{r_1}, s}(t_s)$. These effects become especially pronounced when any stray magnetic fields split up the Zeeman sublevels, which would need to be compensated well below the magnitude of the earth magnetic field for high accuracy. Stray electric fields, if sufficiently strong, would additionally lead to state mixing affecting the underlying physical lifetime of the measured state. According to our calculations (detailed in \cite{Grimmel.2015}) this would become relevant on a \SI{1}{\percent} (probability) level at \SI{30}{\volt\per\centi\metre} for the $30S_{1/2 (|m_j|{=}1/2)}$ state and \SI{2.2}{\volt\per\centi\metre} for $38D_{5/2 (|m_j|{=}1/2)}$, i.e. not leading to significant state mixing in our setup.

Regarding the measurements taken at a lowered environment temperature, mainly two sources of systematic error need to be taken into account: First, as mentioned before, and discussed in \cite{Gallagher.1994}, the effective temperature as seen by the atoms will not be the temperature of the cryostat unless the full $4\pi$ solid angle is covered. Second, the geometry of conducting parts of the experimental setup, like the radiation shield of the cryostat, can form an effective microwave resonator altering the BBR spectrum. This becomes especially relevant if their dimensions are close to the strongest transition wavelengths (like the $30S\leftrightarrow30P$ with a wavelength of \SI{1.9}{\milli\meter}, or $30S\leftrightarrow29P$ with \SI{1.7}{\milli\meter}).

\section{Conclusion and Outlook \label{sec:conclusion}}

We have presented an optical method for the measurement of Rydberg atom lifetimes, providing information about how they are influenced by effects like blackbody radiation and interactions between Rydberg atoms.
Due to the fact that no detector parts or high voltages are needed inside the vacuum chamber of the experimental setup this method might prove particularly useful in otherwise complex systems, including atom chip setups such as the one used here \cite{Cano.2011}, or millikelvin systems \cite{Jessen.2014}, that might eventually lead to the realization of proposed quantum gate schemes as in \cite{Petrosyan.2008} and \cite{Pritchard.2014}.

While the use of an additional microwave population transfer as in \cite{Branden.2010} is not necessary for state selectivity it might be employed to measure the populations of neighboring Rydberg $n P$ and $n F$ states by generalizing the scheme to measure signal differences due to the microwave transfer. This way, both the populations of e.g. a Rydberg $n S$ and close-lying $n P$ states could be monitored. This would, for example, allow for the distinction between superradiance, which highly depends on the population of such strongly coupled states, and other lifetime-reducing effects.

\appendix* 

\begin{acknowledgments}
This work was financially supported by the FET-Open Xtrack Project HAIRS and the Carl Zeiss Stiftung.
\end{acknowledgments}


\begin{thebibliography}{32}%
	\makeatletter
	\providecommand \@ifxundefined [1]{%
		\@ifx{#1\undefined}
	}%
	\providecommand \@ifnum [1]{%
		\ifnum #1\expandafter \@firstoftwo
		\else \expandafter \@secondoftwo
		\fi
	}%
	\providecommand \@ifx [1]{%
		\ifx #1\expandafter \@firstoftwo
		\else \expandafter \@secondoftwo
		\fi
	}%
	\providecommand \natexlab [1]{#1}%
	\providecommand \enquote  [1]{``#1''}%
	\providecommand \bibnamefont  [1]{#1}%
	\providecommand \bibfnamefont [1]{#1}%
	\providecommand \citenamefont [1]{#1}%
	\providecommand \href@noop [0]{\@secondoftwo}%
	\providecommand \href [0]{\begingroup \@sanitize@url \@href}%
	\providecommand \@href[1]{\@@startlink{#1}\@@href}%
	\providecommand \@@href[1]{\endgroup#1\@@endlink}%
	\providecommand \@sanitize@url [0]{\catcode `\\12\catcode `\$12\catcode
		`\&12\catcode `\#12\catcode `\^12\catcode `\_12\catcode `\%12\relax}%
	\providecommand \@@startlink[1]{}%
	\providecommand \@@endlink[0]{}%
	\providecommand \url  [0]{\begingroup\@sanitize@url \@url }%
	\providecommand \@url [1]{\endgroup\@href {#1}{\urlprefix }}%
	\providecommand \urlprefix  [0]{URL }%
	\providecommand \Eprint [0]{\href }%
	\providecommand \doibase [0]{http://dx.doi.org/}%
	\providecommand \selectlanguage [0]{\@gobble}%
	\providecommand \bibinfo  [0]{\@secondoftwo}%
	\providecommand \bibfield  [0]{\@secondoftwo}%
	\providecommand \translation [1]{[#1]}%
	\providecommand \BibitemOpen [0]{}%
	\providecommand \bibitemStop [0]{}%
	\providecommand \bibitemNoStop [0]{.\EOS\space}%
	\providecommand \EOS [0]{\spacefactor3000\relax}%
	\providecommand \BibitemShut  [1]{\csname bibitem#1\endcsname}%
	\let\auto@bib@innerbib\@empty
	\bibitem [{\citenamefont {Saffman}\ \emph {et~al.}(2010)\citenamefont
		{Saffman}, \citenamefont {Walker},\ and\ \citenamefont
		{M\o{}lmer}}]{Saffman.2010}%
	\BibitemOpen
	\bibfield  {author} {\bibinfo {author} {\bibfnamefont {M.}~\bibnamefont
			{Saffman}}, \bibinfo {author} {\bibfnamefont {T.~G.}\ \bibnamefont {Walker}},
		\ and\ \bibinfo {author} {\bibfnamefont {K.}~\bibnamefont {M\o{}lmer}},\
	}\href {\doibase 10.1103/RevModPhys.82.2313} {\bibfield  {journal} {\bibinfo
		{journal} {Rev. Mod. Phys.}\ }\textbf {\bibinfo {volume} {82}},\ \bibinfo
	{pages} {2313} (\bibinfo {year} {2010})}\BibitemShut {NoStop}%
\bibitem [{\citenamefont {Jaksch}\ \emph {et~al.}(2000)\citenamefont {Jaksch},
	\citenamefont {Cirac}, \citenamefont {Zoller}, \citenamefont {Rolston},
	\citenamefont {C\^ot\'e},\ and\ \citenamefont {Lukin}}]{Jaksch.2000}%
\BibitemOpen
\bibfield  {author} {\bibinfo {author} {\bibfnamefont {D.}~\bibnamefont
		{Jaksch}}, \bibinfo {author} {\bibfnamefont {J.~I.}\ \bibnamefont {Cirac}},
	\bibinfo {author} {\bibfnamefont {P.}~\bibnamefont {Zoller}}, \bibinfo
	{author} {\bibfnamefont {S.~L.}\ \bibnamefont {Rolston}}, \bibinfo {author}
	{\bibfnamefont {R.}~\bibnamefont {C\^ot\'e}}, \ and\ \bibinfo {author}
	{\bibfnamefont {M.~D.}\ \bibnamefont {Lukin}},\ }\href {\doibase
	10.1103/PhysRevLett.85.2208} {\bibfield  {journal} {\bibinfo  {journal}
		{Phys. Rev. Lett.}\ }\textbf {\bibinfo {volume} {85}},\ \bibinfo {pages}
	{2208} (\bibinfo {year} {2000})}\BibitemShut {NoStop}%
\bibitem [{\citenamefont {M\"uller}\ \emph {et~al.}(2014)\citenamefont
	{M\"uller}, \citenamefont {Murphy}, \citenamefont {Montangero}, \citenamefont
	{Calarco}, \citenamefont {Grangier},\ and\ \citenamefont
	{Browaeys}}]{Mueller.2014}%
\BibitemOpen
\bibfield  {author} {\bibinfo {author} {\bibfnamefont {M.~M.}\ \bibnamefont
		{M\"uller}}, \bibinfo {author} {\bibfnamefont {M.}~\bibnamefont {Murphy}},
	\bibinfo {author} {\bibfnamefont {S.}~\bibnamefont {Montangero}}, \bibinfo
	{author} {\bibfnamefont {T.}~\bibnamefont {Calarco}}, \bibinfo {author}
	{\bibfnamefont {P.}~\bibnamefont {Grangier}}, \ and\ \bibinfo {author}
	{\bibfnamefont {A.}~\bibnamefont {Browaeys}},\ }\href {\doibase
	10.1103/PhysRevA.89.032334} {\bibfield  {journal} {\bibinfo  {journal} {Phys.
			Rev. A}\ }\textbf {\bibinfo {volume} {89}},\ \bibinfo {pages} {032334}
	(\bibinfo {year} {2014})}\BibitemShut {NoStop}%
\bibitem [{\citenamefont {Beterov}\ \emph
	{et~al.}(2009{\natexlab{a}})\citenamefont {Beterov}, \citenamefont
	{Ryabtsev}, \citenamefont {Tretyakov},\ and\ \citenamefont
	{Entin}}]{Beterov.2009}%
\BibitemOpen
\bibfield  {author} {\bibinfo {author} {\bibfnamefont {I.~I.}\ \bibnamefont
		{Beterov}}, \bibinfo {author} {\bibfnamefont {I.~I.}\ \bibnamefont
		{Ryabtsev}}, \bibinfo {author} {\bibfnamefont {D.~B.}\ \bibnamefont
		{Tretyakov}}, \ and\ \bibinfo {author} {\bibfnamefont {V.~M.}\ \bibnamefont
		{Entin}},\ }\href {\doibase 10.1103/PhysRevA.79.052504} {\bibfield  {journal}
	{\bibinfo  {journal} {Phys. Rev. A}\ }\textbf {\bibinfo {volume} {79}},\
	\bibinfo {pages} {052504} (\bibinfo {year} {2009}{\natexlab{a}})}\BibitemShut
{NoStop}%
\bibitem [{\citenamefont {Beterov}\ \emph
	{et~al.}(2009{\natexlab{b}})\citenamefont {Beterov}, \citenamefont
	{Ryabtsev}, \citenamefont {Tretyakov},\ and\ \citenamefont
	{Entin}}]{Beterov.2009.Err}%
\BibitemOpen
\bibfield  {author} {\bibinfo {author} {\bibfnamefont {I.~I.}\ \bibnamefont
		{Beterov}}, \bibinfo {author} {\bibfnamefont {I.~I.}\ \bibnamefont
		{Ryabtsev}}, \bibinfo {author} {\bibfnamefont {D.~B.}\ \bibnamefont
		{Tretyakov}}, \ and\ \bibinfo {author} {\bibfnamefont {V.~M.}\ \bibnamefont
		{Entin}},\ }\href {\doibase 10.1103/PhysRevA.80.059902} {\bibfield  {journal}
	{\bibinfo  {journal} {Phys. Rev. A}\ }\textbf {\bibinfo {volume} {80}},\
	\bibinfo {pages} {059902(E)} (\bibinfo {year}
	{2009}{\natexlab{b}})}\BibitemShut {NoStop}%
\bibitem [{\citenamefont {Walz-Flannigan}\ \emph {et~al.}(2004)\citenamefont
	{Walz-Flannigan}, \citenamefont {Guest}, \citenamefont {Choi},\ and\
	\citenamefont {Raithel}}]{WalzFlannigan.2004}%
\BibitemOpen
\bibfield  {author} {\bibinfo {author} {\bibfnamefont {A.}~\bibnamefont
		{Walz-Flannigan}}, \bibinfo {author} {\bibfnamefont {J.~R.}\ \bibnamefont
		{Guest}}, \bibinfo {author} {\bibfnamefont {J.-H.}\ \bibnamefont {Choi}}, \
	and\ \bibinfo {author} {\bibfnamefont {G.}~\bibnamefont {Raithel}},\ }\href
{\doibase 10.1103/PhysRevA.69.063405} {\bibfield  {journal} {\bibinfo
		{journal} {Phys. Rev. A}\ }\textbf {\bibinfo {volume} {69}},\ \bibinfo
	{pages} {063405} (\bibinfo {year} {2004})}\BibitemShut {NoStop}%
\bibitem [{\citenamefont {Wang}\ \emph {et~al.}(2007)\citenamefont {Wang},
	\citenamefont {Yelin}, \citenamefont {C\^ot\'e}, \citenamefont {Eyler},
	\citenamefont {Farooqi}, \citenamefont {Gould}, \citenamefont
	{Ko\ifmmode~\check{s}\else \v{s}\fi{}trun}, \citenamefont {Tong},\ and\
	\citenamefont {Vrinceanu}}]{Wang.2007}%
\BibitemOpen
\bibfield  {author} {\bibinfo {author} {\bibfnamefont {T.}~\bibnamefont
		{Wang}}, \bibinfo {author} {\bibfnamefont {S.~F.}\ \bibnamefont {Yelin}},
	\bibinfo {author} {\bibfnamefont {R.}~\bibnamefont {C\^ot\'e}}, \bibinfo
	{author} {\bibfnamefont {E.~E.}\ \bibnamefont {Eyler}}, \bibinfo {author}
	{\bibfnamefont {S.~M.}\ \bibnamefont {Farooqi}}, \bibinfo {author}
	{\bibfnamefont {P.~L.}\ \bibnamefont {Gould}}, \bibinfo {author}
	{\bibfnamefont {M.}~\bibnamefont {Ko\ifmmode~\check{s}\else \v{s}\fi{}trun}},
	\bibinfo {author} {\bibfnamefont {D.}~\bibnamefont {Tong}}, \ and\ \bibinfo
	{author} {\bibfnamefont {D.}~\bibnamefont {Vrinceanu}},\ }\href {\doibase
	10.1103/PhysRevA.75.033802} {\bibfield  {journal} {\bibinfo  {journal} {Phys.
			Rev. A}\ }\textbf {\bibinfo {volume} {75}},\ \bibinfo {pages} {033802}
	(\bibinfo {year} {2007})}\BibitemShut {NoStop}%
\bibitem [{\citenamefont {Gallagher}(1994)}]{Gallagher.1994}%
\BibitemOpen
\bibfield  {author} {\bibinfo {author} {\bibfnamefont {T.~F.}\ \bibnamefont
		{Gallagher}},\ }\href@noop {} {\emph {\bibinfo {title} {Rydberg Atoms}}},\
\bibinfo {edition} {1st}\ ed.\ (\bibinfo  {publisher} {Cambridge Univ.
	Press},\ \bibinfo {address} {Cambridge},\ \bibinfo {year} {1994})\BibitemShut
{NoStop}%
\bibitem [{\citenamefont {Spencer}\ \emph {et~al.}(1982)\citenamefont
	{Spencer}, \citenamefont {Vaidyanathan}, \citenamefont {Kleppner},\ and\
	\citenamefont {Ducas}}]{Spencer.1982}%
\BibitemOpen
\bibfield  {author} {\bibinfo {author} {\bibfnamefont {W.~P.}\ \bibnamefont
		{Spencer}}, \bibinfo {author} {\bibfnamefont {A.~G.}\ \bibnamefont
		{Vaidyanathan}}, \bibinfo {author} {\bibfnamefont {D.}~\bibnamefont
		{Kleppner}}, \ and\ \bibinfo {author} {\bibfnamefont {T.~W.}\ \bibnamefont
		{Ducas}},\ }\href {\doibase 10.1103/PhysRevA.25.380} {\bibfield  {journal}
	{\bibinfo  {journal} {Phys. Rev. A}\ }\textbf {\bibinfo {volume} {25}},\
	\bibinfo {pages} {380} (\bibinfo {year} {1982})}\BibitemShut {NoStop}%
\bibitem [{\citenamefont {Grimmel}\ \emph {et~al.}(2015)\citenamefont
	{Grimmel}, \citenamefont {Mack}, \citenamefont {Karlewski}, \citenamefont
	{Jessen}, \citenamefont {Reinschmidt}, \citenamefont {S\'andor},\ and\
	\citenamefont {Fort\'agh}}]{Grimmel.2015}%
\BibitemOpen
\bibfield  {author} {\bibinfo {author} {\bibfnamefont {J.}~\bibnamefont
		{Grimmel}}, \bibinfo {author} {\bibfnamefont {M.}~\bibnamefont {Mack}},
	\bibinfo {author} {\bibfnamefont {F.}~\bibnamefont {Karlewski}}, \bibinfo
	{author} {\bibfnamefont {F.}~\bibnamefont {Jessen}}, \bibinfo {author}
	{\bibfnamefont {M.}~\bibnamefont {Reinschmidt}}, \bibinfo {author}
	{\bibfnamefont {N.}~\bibnamefont {S\'andor}}, \ and\ \bibinfo {author}
	{\bibfnamefont {J.}~\bibnamefont {Fort\'agh}},\ }\href
{http://stacks.iop.org/1367-2630/17/i=5/a=053005} {\bibfield  {journal}
	{\bibinfo  {journal} {New Journal of Physics}\ }\textbf {\bibinfo {volume}
		{17}},\ \bibinfo {pages} {053005} (\bibinfo {year} {2015})}\BibitemShut
{NoStop}%
\bibitem [{\citenamefont {Beterov}\ \emph
	{et~al.}(2009{\natexlab{c}})\citenamefont {Beterov}, \citenamefont
	{Tretyakov}, \citenamefont {Ryabtsev}, \citenamefont {Entin}, \citenamefont
	{Ekers},\ and\ \citenamefont {Bezuglov}}]{Beterov.2009_2}%
\BibitemOpen
\bibfield  {author} {\bibinfo {author} {\bibfnamefont {I.~I.}\ \bibnamefont
		{Beterov}}, \bibinfo {author} {\bibfnamefont {D.~B.}\ \bibnamefont
		{Tretyakov}}, \bibinfo {author} {\bibfnamefont {I.~I.}\ \bibnamefont
		{Ryabtsev}}, \bibinfo {author} {\bibfnamefont {V.~M.}\ \bibnamefont {Entin}},
	\bibinfo {author} {\bibfnamefont {A.}~\bibnamefont {Ekers}}, \ and\ \bibinfo
	{author} {\bibfnamefont {N.~N.}\ \bibnamefont {Bezuglov}},\ }\href
{http://stacks.iop.org/1367-2630/11/i=1/a=013052} {\bibfield  {journal}
	{\bibinfo  {journal} {New Journal of Physics}\ }\textbf {\bibinfo {volume}
		{11}},\ \bibinfo {pages} {013052} (\bibinfo {year}
	{2009}{\natexlab{c}})}\BibitemShut {NoStop}%
\bibitem [{\citenamefont {Spencer}\ \emph {et~al.}(1981)\citenamefont
	{Spencer}, \citenamefont {Vaidyanathan}, \citenamefont {Kleppner},\ and\
	\citenamefont {Ducas}}]{Spencer.1981}%
\BibitemOpen
\bibfield  {author} {\bibinfo {author} {\bibfnamefont {W.~P.}\ \bibnamefont
		{Spencer}}, \bibinfo {author} {\bibfnamefont {A.~G.}\ \bibnamefont
		{Vaidyanathan}}, \bibinfo {author} {\bibfnamefont {D.}~\bibnamefont
		{Kleppner}}, \ and\ \bibinfo {author} {\bibfnamefont {T.~W.}\ \bibnamefont
		{Ducas}},\ }\href {\doibase 10.1103/PhysRevA.24.2513} {\bibfield  {journal}
	{\bibinfo  {journal} {Phys. Rev. A}\ }\textbf {\bibinfo {volume} {24}},\
	\bibinfo {pages} {2513} (\bibinfo {year} {1981})}\BibitemShut {NoStop}%
\bibitem [{\citenamefont {de~Oliveira}\ \emph {et~al.}(2002)\citenamefont
	{de~Oliveira}, \citenamefont {Mancini}, \citenamefont {Bagnato},\ and\
	\citenamefont {Marcassa}}]{Oliveira.2002}%
\BibitemOpen
\bibfield  {author} {\bibinfo {author} {\bibfnamefont {A.~L.}\ \bibnamefont
		{de~Oliveira}}, \bibinfo {author} {\bibfnamefont {M.~W.}\ \bibnamefont
		{Mancini}}, \bibinfo {author} {\bibfnamefont {V.~S.}\ \bibnamefont
		{Bagnato}}, \ and\ \bibinfo {author} {\bibfnamefont {L.~G.}\ \bibnamefont
		{Marcassa}},\ }\href {\doibase 10.1103/PhysRevA.65.031401} {\bibfield
	{journal} {\bibinfo  {journal} {Phys. Rev. A}\ }\textbf {\bibinfo {volume}
		{65}},\ \bibinfo {pages} {031401} (\bibinfo {year} {2002})}\BibitemShut
{NoStop}%
\bibitem [{\citenamefont {Nascimento}\ \emph {et~al.}(2006)\citenamefont
	{Nascimento}, \citenamefont {Caliri}, \citenamefont {de~Oliveira},
	\citenamefont {Bagnato},\ and\ \citenamefont {Marcassa}}]{Nascimento.2006}%
\BibitemOpen
\bibfield  {author} {\bibinfo {author} {\bibfnamefont {V.~A.}\ \bibnamefont
		{Nascimento}}, \bibinfo {author} {\bibfnamefont {L.~L.}\ \bibnamefont
		{Caliri}}, \bibinfo {author} {\bibfnamefont {A.~L.}\ \bibnamefont
		{de~Oliveira}}, \bibinfo {author} {\bibfnamefont {V.~S.}\ \bibnamefont
		{Bagnato}}, \ and\ \bibinfo {author} {\bibfnamefont {L.~G.}\ \bibnamefont
		{Marcassa}},\ }\href {\doibase 10.1103/PhysRevA.74.054501} {\bibfield
	{journal} {\bibinfo  {journal} {Phys. Rev. A}\ }\textbf {\bibinfo {volume}
		{74}},\ \bibinfo {pages} {054501} (\bibinfo {year} {2006})}\BibitemShut
{NoStop}%
\bibitem [{\citenamefont {Tate}(2007)}]{Tate.2007}%
\BibitemOpen
\bibfield  {author} {\bibinfo {author} {\bibfnamefont {D.~A.}\ \bibnamefont
		{Tate}},\ }\href {\doibase 10.1103/PhysRevA.75.066502} {\bibfield  {journal}
	{\bibinfo  {journal} {Phys. Rev. A}\ }\textbf {\bibinfo {volume} {75}},\
	\bibinfo {pages} {066502} (\bibinfo {year} {2007})}\BibitemShut {NoStop}%
\bibitem [{\citenamefont {Caliri}\ and\ \citenamefont
	{Marcassa}(2007)}]{Caliri.2007}%
\BibitemOpen
\bibfield  {author} {\bibinfo {author} {\bibfnamefont {L.~L.}\ \bibnamefont
		{Caliri}}\ and\ \bibinfo {author} {\bibfnamefont {L.~G.}\ \bibnamefont
		{Marcassa}},\ }\href {\doibase 10.1103/PhysRevA.75.066503} {\bibfield
	{journal} {\bibinfo  {journal} {Phys. Rev. A}\ }\textbf {\bibinfo {volume}
		{75}},\ \bibinfo {pages} {066503} (\bibinfo {year} {2007})}\BibitemShut
{NoStop}%
\bibitem [{\citenamefont {Branden}\ \emph {et~al.}(2010)\citenamefont
	{Branden}, \citenamefont {Juhasz}, \citenamefont {Mahlokozera}, \citenamefont
	{Vesa}, \citenamefont {Wilson}, \citenamefont {Zheng}, \citenamefont
	{Kortyna},\ and\ \citenamefont {Tate}}]{Branden.2010}%
\BibitemOpen
\bibfield  {author} {\bibinfo {author} {\bibfnamefont {D.~B.}\ \bibnamefont
		{Branden}}, \bibinfo {author} {\bibfnamefont {T.}~\bibnamefont {Juhasz}},
	\bibinfo {author} {\bibfnamefont {T.}~\bibnamefont {Mahlokozera}}, \bibinfo
	{author} {\bibfnamefont {C.}~\bibnamefont {Vesa}}, \bibinfo {author}
	{\bibfnamefont {R.~O.}\ \bibnamefont {Wilson}}, \bibinfo {author}
	{\bibfnamefont {M.}~\bibnamefont {Zheng}}, \bibinfo {author} {\bibfnamefont
		{A.}~\bibnamefont {Kortyna}}, \ and\ \bibinfo {author} {\bibfnamefont
		{D.~A.}\ \bibnamefont {Tate}},\ }\href
{http://stacks.iop.org/0953-4075/43/i=1/a=015002} {\bibfield  {journal}
	{\bibinfo  {journal} {J. Phys. B}\ }\textbf {\bibinfo {volume} {43}},\
	\bibinfo {pages} {015002} (\bibinfo {year} {2010})}\BibitemShut {NoStop}%
\bibitem [{\citenamefont {Li}\ \emph {et~al.}(2005)\citenamefont {Li},
	\citenamefont {Tanner},\ and\ \citenamefont {Gallagher}}]{Li.2005}%
\BibitemOpen
\bibfield  {author} {\bibinfo {author} {\bibfnamefont {W.}~\bibnamefont
		{Li}}, \bibinfo {author} {\bibfnamefont {P.~J.}\ \bibnamefont {Tanner}}, \
	and\ \bibinfo {author} {\bibfnamefont {T.~F.}\ \bibnamefont {Gallagher}},\
}\href {\doibase 10.1103/PhysRevLett.94.173001} {\bibfield  {journal}
{\bibinfo  {journal} {Phys. Rev. Lett.}\ }\textbf {\bibinfo {volume} {94}},\
\bibinfo {pages} {173001} (\bibinfo {year} {2005})}\BibitemShut {NoStop}%
\bibitem [{\citenamefont {Zanon}\ \emph {et~al.}(2002)\citenamefont {Zanon},
	\citenamefont {Magalh\~aes}, \citenamefont {de~Oliveira},\ and\ \citenamefont
	{Marcassa}}]{Zanon.2002}%
\BibitemOpen
\bibfield  {author} {\bibinfo {author} {\bibfnamefont {R.~A. D.~S.}\
		\bibnamefont {Zanon}}, \bibinfo {author} {\bibfnamefont {K.~M.~F.}\
		\bibnamefont {Magalh\~aes}}, \bibinfo {author} {\bibfnamefont {A.~L.}\
		\bibnamefont {de~Oliveira}}, \ and\ \bibinfo {author} {\bibfnamefont {L.~G.}\
		\bibnamefont {Marcassa}},\ }\href {\doibase 10.1103/PhysRevA.65.023405}
{\bibfield  {journal} {\bibinfo  {journal} {Phys. Rev. A}\ }\textbf {\bibinfo
		{volume} {65}},\ \bibinfo {pages} {023405} (\bibinfo {year}
	{2002})}\BibitemShut {NoStop}%
\bibitem [{\citenamefont {Day}\ \emph {et~al.}(2008)\citenamefont {Day},
	\citenamefont {Brekke},\ and\ \citenamefont {Walker}}]{Day.2008}%
\BibitemOpen
\bibfield  {author} {\bibinfo {author} {\bibfnamefont {J.~O.}\ \bibnamefont
		{Day}}, \bibinfo {author} {\bibfnamefont {E.}~\bibnamefont {Brekke}}, \ and\
	\bibinfo {author} {\bibfnamefont {T.~G.}\ \bibnamefont {Walker}},\ }\href
{\doibase 10.1103/PhysRevA.77.052712} {\bibfield  {journal} {\bibinfo
		{journal} {Phys. Rev. A}\ }\textbf {\bibinfo {volume} {77}},\ \bibinfo
	{pages} {052712} (\bibinfo {year} {2008})}\BibitemShut {NoStop}%
\bibitem [{\citenamefont {Tauschinsky}\ \emph {et~al.}(2010)\citenamefont
	{Tauschinsky}, \citenamefont {Thijssen}, \citenamefont {Whitlock},
	\citenamefont {van Linden van~den Heuvell},\ and\ \citenamefont
	{Spreeuw}}]{Tauschinsky.2010}%
\BibitemOpen
\bibfield  {author} {\bibinfo {author} {\bibfnamefont {A.}~\bibnamefont
		{Tauschinsky}}, \bibinfo {author} {\bibfnamefont {R.~M.~T.}\ \bibnamefont
		{Thijssen}}, \bibinfo {author} {\bibfnamefont {S.}~\bibnamefont {Whitlock}},
	\bibinfo {author} {\bibfnamefont {H.~B.}\ \bibnamefont {van Linden van~den
			Heuvell}}, \ and\ \bibinfo {author} {\bibfnamefont {R.~J.~C.}\ \bibnamefont
		{Spreeuw}},\ }\href {\doibase 10.1103/PhysRevA.81.063411} {\bibfield
	{journal} {\bibinfo  {journal} {Phys. Rev. A}\ }\textbf {\bibinfo {volume}
		{81}},\ \bibinfo {pages} {063411} (\bibinfo {year} {2010})}\BibitemShut
{NoStop}%
\bibitem [{\citenamefont {G\"unter}\ \emph {et~al.}(2013)\citenamefont
	{G\"unter}, \citenamefont {Schempp}, \citenamefont {Robert-de Saint-Vincent},
	\citenamefont {Gavryusev}, \citenamefont {Helmrich}, \citenamefont {Hofmann},
	\citenamefont {Whitlock},\ and\ \citenamefont
	{Weidem\"uller}}]{Guenter.2013}%
\BibitemOpen
\bibfield  {author} {\bibinfo {author} {\bibfnamefont {G.}~\bibnamefont
		{G\"unter}}, \bibinfo {author} {\bibfnamefont {H.}~\bibnamefont {Schempp}},
	\bibinfo {author} {\bibfnamefont {M.}~\bibnamefont {Robert-de
			Saint-Vincent}}, \bibinfo {author} {\bibfnamefont {V.}~\bibnamefont
		{Gavryusev}}, \bibinfo {author} {\bibfnamefont {S.}~\bibnamefont {Helmrich}},
	\bibinfo {author} {\bibfnamefont {C.~S.}\ \bibnamefont {Hofmann}}, \bibinfo
	{author} {\bibfnamefont {S.}~\bibnamefont {Whitlock}}, \ and\ \bibinfo
	{author} {\bibfnamefont {M.}~\bibnamefont {Weidem\"uller}},\ }\href {\doibase
	10.1126/science.1244843} {\bibfield  {journal} {\bibinfo  {journal}
		{Science}\ }\textbf {\bibinfo {volume} {342}},\ \bibinfo {pages} {954}
	(\bibinfo {year} {2013})}\BibitemShut {NoStop}%
\bibitem [{\citenamefont {Maxwell}\ \emph {et~al.}(2014)\citenamefont
	{Maxwell}, \citenamefont {Szwer}, \citenamefont {Paredes-Barato},
	\citenamefont {Busche}, \citenamefont {Pritchard}, \citenamefont {Gauguet},
	\citenamefont {Jones},\ and\ \citenamefont {Adams}}]{Maxwell.2014}%
\BibitemOpen
\bibfield  {author} {\bibinfo {author} {\bibfnamefont {D.}~\bibnamefont
		{Maxwell}}, \bibinfo {author} {\bibfnamefont {D.~J.}\ \bibnamefont {Szwer}},
	\bibinfo {author} {\bibfnamefont {D.}~\bibnamefont {Paredes-Barato}},
	\bibinfo {author} {\bibfnamefont {H.}~\bibnamefont {Busche}}, \bibinfo
	{author} {\bibfnamefont {J.~D.}\ \bibnamefont {Pritchard}}, \bibinfo {author}
	{\bibfnamefont {A.}~\bibnamefont {Gauguet}}, \bibinfo {author} {\bibfnamefont
		{M.~P.~A.}\ \bibnamefont {Jones}}, \ and\ \bibinfo {author} {\bibfnamefont
		{C.~S.}\ \bibnamefont {Adams}},\ }\href {\doibase 10.1103/PhysRevA.89.043827}
{\bibfield  {journal} {\bibinfo  {journal} {Phys. Rev. A}\ }\textbf {\bibinfo
		{volume} {89}},\ \bibinfo {pages} {043827} (\bibinfo {year}
	{2014})}\BibitemShut {NoStop}%
\bibitem [{\citenamefont {Gorniaczyk}\ \emph {et~al.}(2014)\citenamefont
	{Gorniaczyk}, \citenamefont {Tresp}, \citenamefont {Schmidt}, \citenamefont
	{Fedder},\ and\ \citenamefont {Hofferberth}}]{Gorniaczyk.2014}%
\BibitemOpen
\bibfield  {author} {\bibinfo {author} {\bibfnamefont {H.}~\bibnamefont
		{Gorniaczyk}}, \bibinfo {author} {\bibfnamefont {C.}~\bibnamefont {Tresp}},
	\bibinfo {author} {\bibfnamefont {J.}~\bibnamefont {Schmidt}}, \bibinfo
	{author} {\bibfnamefont {H.}~\bibnamefont {Fedder}}, \ and\ \bibinfo {author}
	{\bibfnamefont {S.}~\bibnamefont {Hofferberth}},\ }\href {\doibase
	10.1103/PhysRevLett.113.053601} {\bibfield  {journal} {\bibinfo  {journal}
		{Phys. Rev. Lett.}\ }\textbf {\bibinfo {volume} {113}},\ \bibinfo {pages}
	{053601} (\bibinfo {year} {2014})}\BibitemShut {NoStop}%
\bibitem [{\citenamefont {Karlewski}\ \emph {et~al.}(2015)\citenamefont
	{Karlewski}, \citenamefont {Mack}, \citenamefont {Grimmel}, \citenamefont
	{S\'andor},\ and\ \citenamefont {Fort\'agh}}]{Karlewski.2015}%
\BibitemOpen
\bibfield  {author} {\bibinfo {author} {\bibfnamefont {F.}~\bibnamefont
		{Karlewski}}, \bibinfo {author} {\bibfnamefont {M.}~\bibnamefont {Mack}},
	\bibinfo {author} {\bibfnamefont {J.}~\bibnamefont {Grimmel}}, \bibinfo
	{author} {\bibfnamefont {N.}~\bibnamefont {S\'andor}}, \ and\ \bibinfo
	{author} {\bibfnamefont {J.}~\bibnamefont {Fort\'agh}},\ }\href {\doibase
	10.1103/PhysRevA.91.043422} {\bibfield  {journal} {\bibinfo  {journal} {Phys.
			Rev. A}\ }\textbf {\bibinfo {volume} {91}},\ \bibinfo {pages} {043422}
	(\bibinfo {year} {2015})}\BibitemShut {NoStop}%
\bibitem [{\citenamefont {Cubel}\ \emph {et~al.}(2005)\citenamefont {Cubel},
	\citenamefont {Teo}, \citenamefont {Malinovsky}, \citenamefont {Guest},
	\citenamefont {Reinhard}, \citenamefont {Knuffman}, \citenamefont {Berman},\
	and\ \citenamefont {Raithel}}]{Cubel.2005}%
\BibitemOpen
\bibfield  {author} {\bibinfo {author} {\bibfnamefont {T.}~\bibnamefont
		{Cubel}}, \bibinfo {author} {\bibfnamefont {B.~K.}\ \bibnamefont {Teo}},
	\bibinfo {author} {\bibfnamefont {V.~S.}\ \bibnamefont {Malinovsky}},
	\bibinfo {author} {\bibfnamefont {J.~R.}\ \bibnamefont {Guest}}, \bibinfo
	{author} {\bibfnamefont {A.}~\bibnamefont {Reinhard}}, \bibinfo {author}
	{\bibfnamefont {B.}~\bibnamefont {Knuffman}}, \bibinfo {author}
	{\bibfnamefont {P.~R.}\ \bibnamefont {Berman}}, \ and\ \bibinfo {author}
	{\bibfnamefont {G.}~\bibnamefont {Raithel}},\ }\href {\doibase
	10.1103/PhysRevA.72.023405} {\bibfield  {journal} {\bibinfo  {journal} {Phys.
			Rev. A}\ }\textbf {\bibinfo {volume} {72}},\ \bibinfo {pages} {023405}
	(\bibinfo {year} {2005})}\BibitemShut {NoStop}%
\bibitem [{\citenamefont {{Cano, D.}}\ \emph {et~al.}(2011)\citenamefont
	{{Cano, D.}}, \citenamefont {{Hattermann, H.}}, \citenamefont {{Kasch, B.}},
	\citenamefont {{Zimmermann, C.}}, \citenamefont {{Kleiner, R.}},
	\citenamefont {{Koelle, D.}},\ and\ \citenamefont {{Fortágh,
			J.}}}]{Cano.2011}%
\BibitemOpen
\bibfield  {author} {\bibinfo {author} {\bibnamefont {{Cano, D.}}}, \bibinfo
	{author} {\bibnamefont {{Hattermann, H.}}}, \bibinfo {author} {\bibnamefont
		{{Kasch, B.}}}, \bibinfo {author} {\bibnamefont {{Zimmermann, C.}}}, \bibinfo
	{author} {\bibnamefont {{Kleiner, R.}}}, \bibinfo {author} {\bibnamefont
		{{Koelle, D.}}}, \ and\ \bibinfo {author} {\bibnamefont {{Fortágh, J.}}},\
}\href {\doibase 10.1140/epjd/e2011-10680-8} {\bibfield  {journal} {\bibinfo
	{journal} {Eur. Phys. J. D}\ }\textbf {\bibinfo {volume} {63}},\ \bibinfo
{pages} {17} (\bibinfo {year} {2011})}\BibitemShut {NoStop}%
\bibitem [{\citenamefont {Hattermann}\ \emph {et~al.}(2012)\citenamefont
	{Hattermann}, \citenamefont {Mack}, \citenamefont {Karlewski}, \citenamefont
	{Jessen}, \citenamefont {Cano},\ and\ \citenamefont
	{Fort\'agh}}]{Hattermann.2012}%
\BibitemOpen
\bibfield  {author} {\bibinfo {author} {\bibfnamefont {H.}~\bibnamefont
		{Hattermann}}, \bibinfo {author} {\bibfnamefont {M.}~\bibnamefont {Mack}},
	\bibinfo {author} {\bibfnamefont {F.}~\bibnamefont {Karlewski}}, \bibinfo
	{author} {\bibfnamefont {F.}~\bibnamefont {Jessen}}, \bibinfo {author}
	{\bibfnamefont {D.}~\bibnamefont {Cano}}, \ and\ \bibinfo {author}
	{\bibfnamefont {J.}~\bibnamefont {Fort\'agh}},\ }\href {\doibase
	10.1103/PhysRevA.86.022511} {\bibfield  {journal} {\bibinfo  {journal} {Phys.
			Rev. A}\ }\textbf {\bibinfo {volume} {86}},\ \bibinfo {pages} {022511}
	(\bibinfo {year} {2012})}\BibitemShut {NoStop}%
\bibitem [{\citenamefont {Reinhard}\ \emph {et~al.}(2007)\citenamefont
	{Reinhard}, \citenamefont {Liebisch}, \citenamefont {Knuffman},\ and\
	\citenamefont {Raithel}}]{Reinhard.2007}%
\BibitemOpen
\bibfield  {author} {\bibinfo {author} {\bibfnamefont {A.}~\bibnamefont
		{Reinhard}}, \bibinfo {author} {\bibfnamefont {T.~C.}\ \bibnamefont
		{Liebisch}}, \bibinfo {author} {\bibfnamefont {B.}~\bibnamefont {Knuffman}},
	\ and\ \bibinfo {author} {\bibfnamefont {G.}~\bibnamefont {Raithel}},\ }\href
{\doibase 10.1103/PhysRevA.75.032712} {\bibfield  {journal} {\bibinfo
		{journal} {Phys. Rev. A}\ }\textbf {\bibinfo {volume} {75}},\ \bibinfo
	{pages} {032712} (\bibinfo {year} {2007})}\BibitemShut {NoStop}%
\bibitem [{\citenamefont {Jessen}\ \emph {et~al.}(2014)\citenamefont {Jessen},
	\citenamefont {Knufinke}, \citenamefont {Bell}, \citenamefont {Vergien},
	\citenamefont {Hattermann}, \citenamefont {Weiss}, \citenamefont {Rudolph},
	\citenamefont {Reinschmidt}, \citenamefont {Meyer}, \citenamefont {Gaber},
	\citenamefont {Cano}, \citenamefont {G\"unther}, \citenamefont {Bernon},
	\citenamefont {Koelle}, \citenamefont {Kleiner},\ and\ \citenamefont
	{Fort\'agh}}]{Jessen.2014}%
\BibitemOpen
\bibfield  {author} {\bibinfo {author} {\bibfnamefont {F.}~\bibnamefont
		{Jessen}}, \bibinfo {author} {\bibfnamefont {M.}~\bibnamefont {Knufinke}},
	\bibinfo {author} {\bibfnamefont {S.}~\bibnamefont {Bell}}, \bibinfo {author}
	{\bibfnamefont {P.}~\bibnamefont {Vergien}}, \bibinfo {author} {\bibfnamefont
		{H.}~\bibnamefont {Hattermann}}, \bibinfo {author} {\bibfnamefont
		{P.}~\bibnamefont {Weiss}}, \bibinfo {author} {\bibfnamefont
		{M.}~\bibnamefont {Rudolph}}, \bibinfo {author} {\bibfnamefont
		{M.}~\bibnamefont {Reinschmidt}}, \bibinfo {author} {\bibfnamefont
		{K.}~\bibnamefont {Meyer}}, \bibinfo {author} {\bibfnamefont
		{T.}~\bibnamefont {Gaber}}, \bibinfo {author} {\bibfnamefont
		{D.}~\bibnamefont {Cano}}, \bibinfo {author} {\bibfnamefont {A.}~\bibnamefont
		{G\"unther}}, \bibinfo {author} {\bibfnamefont {S.}~\bibnamefont {Bernon}},
	\bibinfo {author} {\bibfnamefont {D.}~\bibnamefont {Koelle}}, \bibinfo
	{author} {\bibfnamefont {R.}~\bibnamefont {Kleiner}}, \ and\ \bibinfo
	{author} {\bibfnamefont {J.}~\bibnamefont {Fort\'agh}},\ }\href {\doibase
	10.1007/s00340-013-5750-5} {\bibfield  {journal} {\bibinfo  {journal}
		{Applied Physics B}\ }\textbf {\bibinfo {volume} {116}},\ \bibinfo {pages}
	{665} (\bibinfo {year} {2014})}\BibitemShut {NoStop}%
\bibitem [{\citenamefont {Petrosyan}\ and\ \citenamefont
	{Fleischhauer}(2008)}]{Petrosyan.2008}%
\BibitemOpen
\bibfield  {author} {\bibinfo {author} {\bibfnamefont {D.}~\bibnamefont
		{Petrosyan}}\ and\ \bibinfo {author} {\bibfnamefont {M.}~\bibnamefont
		{Fleischhauer}},\ }\href {\doibase 10.1103/PhysRevLett.100.170501} {\bibfield
	{journal} {\bibinfo  {journal} {Phys. Rev. Lett.}\ }\textbf {\bibinfo
		{volume} {100}},\ \bibinfo {pages} {170501} (\bibinfo {year}
	{2008})}\BibitemShut {NoStop}%
\bibitem [{\citenamefont {Pritchard}\ \emph {et~al.}(2014)\citenamefont
	{Pritchard}, \citenamefont {Isaacs}, \citenamefont {Beck}, \citenamefont
	{McDermott},\ and\ \citenamefont {Saffman}}]{Pritchard.2014}%
\BibitemOpen
\bibfield  {author} {\bibinfo {author} {\bibfnamefont {J.~D.}\ \bibnamefont
		{Pritchard}}, \bibinfo {author} {\bibfnamefont {J.~A.}\ \bibnamefont
		{Isaacs}}, \bibinfo {author} {\bibfnamefont {M.~A.}\ \bibnamefont {Beck}},
	\bibinfo {author} {\bibfnamefont {R.}~\bibnamefont {McDermott}}, \ and\
	\bibinfo {author} {\bibfnamefont {M.}~\bibnamefont {Saffman}},\ }\href
{\doibase 10.1103/PhysRevA.89.010301} {\bibfield  {journal} {\bibinfo
		{journal} {Phys. Rev. A}\ }\textbf {\bibinfo {volume} {89}},\ \bibinfo
	{pages} {010301} (\bibinfo {year} {2014})}\BibitemShut {NoStop}%
\end{thebibliography}
\end{document}